\documentclass[aps,twocolumn,showpacs, showkeys]{revtex4}

\usepackage{graphicx}
\usepackage{dcolumn}
\usepackage{bm}

\input{epsf.sty}

\begin{document}
\newcommand\slco{Sr$_{0.9}$La$_{0.1}$CuO$_{2}$}
\newcommand\sgco{Sr$_{0.9}$Gd$_{0.1}$CuO$_{2}$}
\newcommand\smco{Sr$_{0.9}$M$_{0.1}$CuO$_{2}$}
\newcommand\tc{$T_{c}$}
\newcommand\cuo{CuO$_{2}$}
\newcommand\hightc{high-$T_{c}$}
\newcommand\hc{$H_{c2}$}
\newcommand\hcc{$H_{c2}^c$}
\newcommand\hcab{$H_{c2}^{ab}$}
\newcommand\hkink{$H_{\rm kink}$}
\newcommand\hirrab{$H_{\rm irr}^{ab}$}
\newcommand\hirrc{$H_{\rm irr}^{c}$}
\newcommand\hirr{$H_{\rm irr}$}
\newcommand\YBCO{YBa$_{2}$Cu$_{3}$O$_{7-\delta}$}
\newcommand\RBCO{REBa$_{2}$Cu$_{3}$O$_{7-\delta}$}
\newcommand\onelayer{Sr$_{2-x}$La$_{x}$CuO$_{4}$}
\newcommand\PCCO{Pr$_{2-x}$Ce$_{x}$CuO$_{4}$}
\newcommand\BSCCO{Bi$_{2}$Sr$_{2}$CaCu$_{2}$O$_{8-\delta}$}

\title{Dimensionality of superconductivity in the infinite-layer high-temperature cuprate Sr$_{0.9}$M$_{0.1}$CuO$_2$ (M = La, Gd)}
\author{V. S. Zapf,$^1$ N.-C. Yeh,$^1$ A. D. Beyer,$^1$ C. R. Hughes,$^1$ C. H. Mielke,$^2$ N. Harrison,$^2$ M. S. Park,$^3$ K. H. Kim,$^3$ S.-I. Lee,$^3$}

\affiliation{$^1$Department of Physics, California Institute of
Technology, Pasadena, CA \\ $^2$National High Magnetic Field
Laboratory, Los Alamos, NM \\ $^3$Department of Physics, Pohang
University of Science and Technology, Pohang, Korea}

\date{\today}

\begin{abstract}
The high magnetic field phase diagram of the electron-doped
infinite layer high-temperature superconducting (\hightc{})
compound \slco{} was probed by means of penetration depth and
magnetization measurements in pulsed fields to 60 T. An anisotropy
ratio of 8 was detected for the upper critical fields with $H$
parallel (\hcab{}) and perpendicular (\hcc{}) to the \cuo{}
planes, with \hcab{} extrapolating to near the Pauli paramagnetic
limit of 160 T. The longer superconducting coherence length than
the lattice constant along the c-axis indicates that the orbital
degrees of freedom of the pairing wavefunction are three
dimensional. By contrast, low-field magnetization and specific
heat measurements of \sgco{} indicate a coexistence of bulk s-wave
superconductivity with large moment Gd paramagnetism close to the
\cuo{} planes, suggesting a strong confinement of the spin degrees
of freedom of the Cooper pair to the \cuo{} planes. The region
between \hcab{} and the irreversibility line in the magnetization,
\hirrab{}, is anomalously large for an electron-doped \hightc{}
cuprate, suggesting the existence of additional quantum
fluctuations perhaps due to a competing spin-density wave order.
\end{abstract}


 \maketitle

In the \hightc{} cuprate superconductors, anisotropy has been
suggested to play an important role in the superconducting pairing
mechanism and the elevated \tc{} in both experimental and
theoretical work \cite{Anderson87,MapleBook}. It is surprising
therefore to find superconductivity (SC) with $T_c = 43$ K in the
optimal electron-doped infinite-layer cuprates \smco{} (M = La,
Gd), which exhibit only a 16\% difference between the $a$ and $c$
tetragonal lattice parameters. The structure of \smco{} is the
most basic among all \hightc{} cuprates, consisting entirely of
CuO$_2$ sheets separated by rare-earth (RE) ions with tetragonal
lattice parameters $c = 3.41$ \AA{} and $a = 3.95$
\AA{}.\cite{Jung02a} The recent success in producing high-quality
polycrystalline samples of the infinite-layer cuprates with no
observable impurity phases \cite{Jung02a} has engendered a renewed
interest in these compounds. X-ray near-edge absorption
spectroscopy indicate electron doping, \cite{Liu01} and bulk SC
has been verified by powdered magnetization ($M$) measurements
\cite{Kim02a} and specific heat ($C$) measurements (data presented
later in this work). Several recent studies of these high purity
polycrystalline samples suggest three-dimensional (3D)
superconductivity in \slco{}. Scanning tunnel spectroscopy (STS)
measurements \cite{Chen02} indicate an unconventional but
isotropic s-wave superconducting gap with no pseudogap at zero
field. The s-wave symmetry of the gap is also supported by
specific heat measurements and Cu-site substitution
studies,\cite{Jung02b,Chen02,Liu03} although it may be
contradicted by NMR measurements.\cite{Williams02} Kim et al
\cite{Kim02a} estimated the c-axis coherence length ($\xi_c$) from
a Hao-Clem analysis \cite{Hao91} of the reversible magnetization
of grain-aligned polycrystal, and found that $\xi_c$ exceeds the
spacing between the CuO$_2$ planes, indicating 3D
superconductivity. On the other hand, they also find significant
anisotropy between magnetic fields $H \le 5$ T oriented parallel
and perpendicular to the \cuo{} planes, with an anisotropy ratio
$\gamma = \xi_c/\xi_{ab} = H_{c2}^{ab}/H_{c2}^c = 9.3$, which is
larger than $\gamma = 5$ observed in \YBCO\, although much smaller
than $\gamma = 55$ observed in optimally doped \BSCCO{}.
\cite{Kim02a,Kim02b,Farrell89, Farrell88} It is interesting to
note that the only major crystallographic difference between the
a-b and the c directions in \slco{} is the presence of oxygen in
the a-b plane, which allows coupling of adjacent Cu spins and has
been implicated as the cause of antiferromagntic ordering or spin
fluctuations in other members of the \hightc{} cuprate family, as
well as a possible mechanism for superconducting pairing.  The
importance of the \cuo\ planes to the SC in \smco{} is further
supported by the fact that Ni substitution on the Cu site rapidly
suppresses \tc{} whereas out-of-plane Gd substitution on the Sr
site leaves \tc{} unchanged. \cite{Jung02b,Jung03}

In this work we determine the upper critical field \hc{} and the
irreversibility field \hirr{} of \slco{} by means of magnetization
and penetration depth measurements in pulsed magnetic fields up to
60 T in order to directly investigate the degree of upper critical
field anisotropy and the role of vortex fluctuations. We also
present specific heat ($C$) and magnetization ($M$) measurements
in low DC fields to 6 T as a function of temperature ($T$) of
\sgco{}, confirming the bulk coexistence of Gd paramagnetism (PM)
and SC. Our results suggest strong confinement of the spin pairing
wave function to the \cuo{} planes and significant field-induced
superconducting fluctuations.

Noncrystalline samples of \slco{} and \sgco{} were prepared under
high pressures as described previously. \cite{Jung02a}
Magnetization measurements in pulsed magnetic fields were
performed at the National High Magnetic Field Laboratory (NHMFL)
in Los Alamos, NM in a $^3$He refrigerator in a 50 T magnet using
a compensated coil. The sample consisted of four pieces of
polycrystalline \slco{} with a total mass of 4.8 mg to maximize
signal and minimize heating. The irreversibility field \hirr{} was
identified from the onset of reversibility in the $M(H)$ loops.
The penetration depth of \slco{} was determined by measuring the
frequency shift $\Delta f$ of a tunnel diode oscillator (TDO)
resonant tank circuit with the sample contained in one of the
component inductors. \cite{Mielke01} A ten turn 0.7 mm diameter
aluminum coil was tightly wound around the sample with a filling
factor of greater than 90\%, with the coil axis oriented
perpendicular to the pulsed field. To maintain temperature
stability, the sample was thermally anchored to a sapphire plate
and placed in $^3$He exchange gas. Small changes in the resonant
frequency can be related to changes in the penetration depth
$\Delta \lambda$  by $\Delta \lambda =
-\frac{R^2}{r_s}\frac{\Delta f}{f_0}$, where $R$ is the radius of
the coil and $r_s$ is the radius of the sample. \cite{Mielke01} In
our case, $R \sim r_s = 0.7$ mm and the reference frequency $f_0
\sim 60$ MHz such that $\Delta f$ = (0.16 MHz/$\mu$m)$\Delta
\lambda$.

\epsfxsize=210pt
\begin{figure}[tbp]
   \centering
\epsfbox{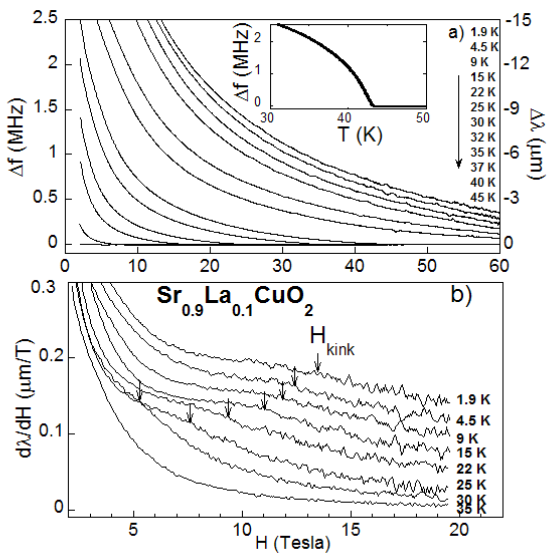} \caption{a) Change in resonant
frequency $\Delta f$ of the TDO tank circuit relative to the
normal state of \slco{} as a function of magnetic field $H$ of
\slco{} polycrystal at various temperatures $T$. The estimated
change in penetration depth $\lambda$ is indicated on the right
axis. Inset: $\Delta f$ as a function of $T$ at zero field where
$T_c = 43 K$. b) Derivative of $\lambda$ with $H$ for various $T$,
with arrows indicating \hkink{}. } \label{penetrationdepth}
\end{figure}

\begin{figure}[tbp]
   \centering
\epsfbox{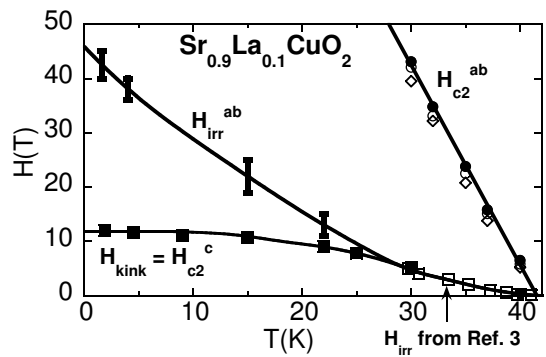} \caption{Field $H$ vs temperature $T$
phase diagram of \slco{}. \hcab{}: onset of SC where $\Delta f$
exceeds 5 kHz in penetration depth measurements. Open circles and
open diamonds indicate the 10 kHz and 20 kHz onsets, respectively.
\hirrab{}: onset of irreversibility in pulsed field $M$ vs $H$
measurements. \hkink{}: change in slope of $\lambda(H)$,
identified with \hcc{} (see text). Open squares: onset of
irreversibility in $M$ vs $T$ measurements by Jung et al.
\cite{Jung02a} Lines are guides to the eye.} \label{phasediagram}
\end{figure}

The frequency shift $\Delta f$ relative to the normal state and
the corresponding $\Delta \lambda$ of \slco{} are shown in Fig.
\ref{penetrationdepth}a as a function of $H$. The inset shows the
$T$-dependence of $\Delta f$ in zero magnetic field. The normal
state resonant frequency $f$ that is reached with increasing field
can only be determined for $T \ge 30$ K; for lower $T$ the sample
remains superconducting to 60 T so $\Delta f$ is estimated. The
frequency shift of the empty coil has been subtracted from all
data. In applied fields, the SC transition is very broad, which
can be attributed to the large anisotropy in \hc{} of the randomly
orientated grains in the polycrystal.\cite{Kim02a} By contrast,
the SC transition as a function of $T$ at $H = 0$ is very sharp,
indicating a high quality sample. Therefore, the onset of
diamagnetism with decreasing $H$ in the $\lambda(H)$ data can be
identified with the largest \hc{}, \hcab{} for fields in the
\cuo{} planes. The onset is defined as $H$ where $\Delta f > 5$
kHz ($\Delta \lambda
> 30$ nm), just above the noise of the experiment. Different onset
criteria have only minor effects on the determination of \hcab{},
as shown in Fig. \ref{phasediagram} where \hcab{} using an onset
criteria of 10 kHz and 20 kHz are shown as open circles and
diamonds, respectively. The upper critical field \hcab{} is linear
in $T$ up to the $H = 60$ T maximum of the experiment, and
extrapolates to 153 T at zero temperature. (Although if we assume
the typical Werthamer-Helfand-Hohenberg curvature for an orbitally
limited superconductor, \cite{Werthamer66} 153 T would constitute
an upper limit for \hcab{}). The linearly extrapolated value of
153 T is close to the s-wave Pauli paramagnetic limit of $H_{c2}^p
= \frac{\Delta}{\sqrt{2}\mu_B}=159$ T, where $\Delta = 13$ meV has
been determined independently from STS data. \cite{Chen02} This
raises the possibility of spin-limited superconductivity for $H$
in the plane, which has also been observed in
\YBCO{}.\cite{OBrien02}

\epsfxsize=210pt

 Determination of the critical fields for $H$ along the c-axis,
\hcc{}, from these data on noncrystalline samples is more
difficult. However, at fields below the onset of diamagnetism we
do observe a significant change in slope of the $\lambda(H)$ data,
indicated as \hkink{} in Fig. \ref{penetrationdepth}b. The
\hkink{} vs $T$ curve is shown in Fig. \ref{phasediagram} and
extrapolates to 12 T at zero temperature. This value of $H_{\rm
kink}(T=0)$ is close to \hcc{} $= 14$ T determined from a Hao-Clem
analysis mentioned previously. \cite{Kim02a} We therefore
associate \hkink{} with \hcc{}. In $\lambda(H)$ measurements of a
noncrystalline sample, a change in slope near \hcc{} could be
expected since the number of grains in the polycrystal that are
superconducting varies with $H$ for \hcc{} $< H < $ \hcab{},
whereas for $H <$ \hcc{} the entire sample is superconducting,
yielding different $H$ dependencies of the flux expulsion in these
two regions. Our data yield an anisotropy ratio $\gamma =
\frac{dH_{c2}^{ab}}{dT}/\frac{H_{c2}^{c}}{dT} \sim 8$, roughly in
agreement with $\gamma = 9.3$ determined from low field studies.
\cite{Kim02a}

In penetration depth measurements of single crystalline organic
superconductors for $H$ along the conducting planes, a kink below
\hc{} for fields has been associated with the vortex melting
transition.\cite{Mielke01} However, in this work we have
determined the vortex dynamics separately by means of
magnetization measurements in pulsed fields, and we find a
significant difference between the onset of irreversibility in
$M(H)$, \hirr{}, and \hkink{} as shown in Fig. \ref{phasediagram}.
Following similar arguments for the $\lambda(H)$ measurements, we
note that \hirrab{} $>$ \hirrc{} in cuprate superconductors,
therefore we assign the onset of irreversibility for
polycrystalline \slco{} to \hirrab{}.  Although we can't rule out
the possibility that \hkink{} might be associated with a vortex
phase transformation, the fact that $H_{\rm kink}(T \rightarrow
0)$ saturates, and $H_{\rm kink}(T \rightarrow 0) << H_{c2}^{ab}(T
\rightarrow 0)$ indicates that $H_{\rm kink}(T)$ is unlikely
caused by a thermally-induced vortex melting transition for $H ||
ab$. Future pulsed-field measurements on grain-aligned or
epitaxial thin film samples will be necessary to conclusively
determine whether \hkink{}$(T)$ obtained in this work may be
identified with \hcc{}$(T)$.

The region between \hirrab{} and \hcab{} in the phase diagram in
Fig. \ref{phasediagram} is significantly larger than is observed
in other electron-doped \hightc{} compounds where \hc{} typically
tracks \hirr{}. \cite{Fournier03,MapleBook} It is particularly
surprising that \hirrab{}$(T \rightarrow 0) \sim 45$ T is much
smaller than \hcab{}$(T \rightarrow 0) \sim 150$ T. In hole-doped
cuprates, a large separation between \hirr{} and \hc{} is often
observed and is generally referred to as a vortex-liquid phase due
to thermally induced fluctuations. The large discrepancy between
\hirrab{} and \hcab{} in e-doped \slco{} even to $T \rightarrow 0$
suggests the presence of field-enhanced SC fluctuations in
\slco{}. Enhanced SC fluctuations down to very low $T$ may be
consistent with a scenario of SC coexisting with a competing
order, such as a spin-density wave (SDW) near a quantum critical
point. \cite{Demler01} In particular, antiferromagnetic spin
fluctuations associated with the competing SDW can be enhanced by
external fields. \cite{Demler01,Sachdev02,Chen03} The conjecture
of a competing order in the SC state of \slco{} is also consistent
with our recent STS studies, \cite{Chen04,Yeh04} where we observe
the emergence of a second energy gap with increasing tunnelling
current upon the closing of the SC gap. In contrast to the SC gap,
the current-induced gap is not spatially uniform probably due to
interactions of the SDW with charge disorder. \cite{Chen03} We
note that experimental evidence for coexistence of a SDW with
cuprate superconductivity has been found in other \hightc
compounds. \cite{Lee99,Kang03,Matsuura03,Lake02,Mook02}

\epsfxsize=210pt

\begin{figure}[tbp]
   \centering
\epsfbox{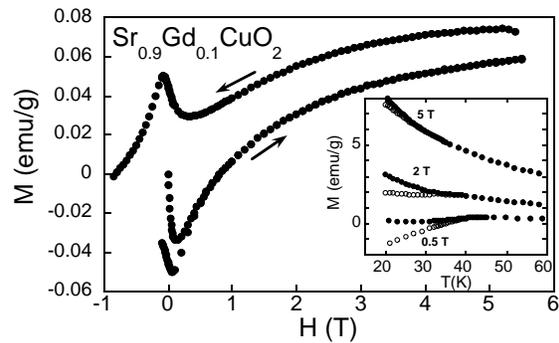}

\caption{Magnetization $M$ vs $H$ of polycrystalline \sgco{} at $T
= 5$ K. Inset shows field-cooled (closed symbols) and
zero-field-cooled (open symbols) $M$ vs $T$ data at $H = 5$ T, 2
T, and 0.5 T.}  \label{Coexistence}
\end{figure}

To further investigate the dimensionality of the
superconductivity, low-field $C(T)$ and $M(T,H)$ of
polycrystalline \sgco{} were measured with $H$ up to 6 T and $T$
down to 1.8 K in a Quantum Design Physical Properties Measurement
System and a SQUID magnetometer, respectively. In contrast to
in-plane Ni substitution on the Cu site, out of plane Gd
substitution on the Sr site does not suppress \tc{}. In fact, the
Gd ions exhibit local moment paramagnetism (PM) that coexists with
SC to low $T$. \cite{Jung03} Figure \ref{Coexistence} shows $M(H)$
at $T = 5$ K, and zero field/field cooled magnetization curves as
a function of $T$ in the inset. The $M(H)$ curve in the main
figure can be viewed as a superposition of a SC hysteresis curve
and a Brillouin function resulting from the PM of Gd. Further
proof for this coexistence is evident in the inset, which shows a
large positive magnetization associated with Gd paramagnetism, but
nevertheless significant hysteresis between zero-field-cooled and
field-cooled curves, indicating superconductivity.

\begin{figure}
   \centering
\epsfbox{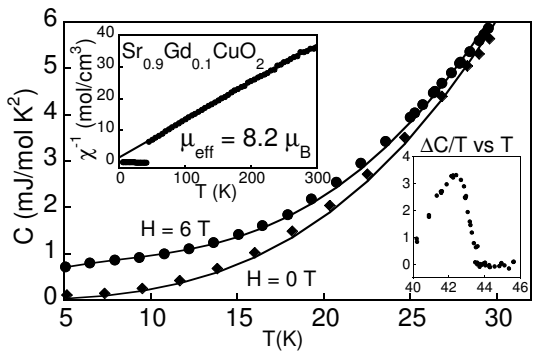} \caption{Main figure: specific heat
$C$ vs $T$ at $H = 0$ and 6 T. Lines are fits to a
field-independent $T^3$ phonon term plus a contribution from Gd
paramagetism, assuming a Gd total momentum of $J = 7/2$. Upper
inset: inverse magnetic susceptibility $1/\chi$ vs $T$ with $H =
100$ Oe, fit by a Curie-Weiss law above $T_c = 43$ K with
$\mu_{\rm eff} = 8.2 \mu_{B}$ and $\theta_{CW} = -27$ K. Lower
inset: electronic specific heat $\Delta C$ vs $T$ showing the
superconducting transition. } \label{SrGdCuO2}
\end{figure}

In Fig. \ref{SrGdCuO2}, the magnitude of the paramagnetic
contribution from the Gd ions is investigated quantitatively. The
main figure shows $C(T)$ at $H = 0$ and $H = 6$ T. The $H = 0$
data is fit by a $T^3$ dependence to model the phonon
contribution, (the electronic contribution at these temperature
can be neglected). For $H = 6$ T, $C(T)$ can be fit by the same
$T^3$ dependence as the $H = 0$ data, plus an additional
contribution from the Gd paramagnetic moments, derived from mean
field theory assuming the Hund's rule $J=7/2$ moment, and one Gd
ion for every ten unit cells. The fit is remarkable, considering
that there are no fitting parameters. In the upper inset of Fig.
\ref{SrGdCuO2}, $1/\chi$ is plotted as function of $T$, where the
line is a Curie-Weiss fit with $\mu_{\rm eff} = 8.2 \mu_B$, which
is close to the Hund's rule moment of 7.6 $\mu_B$ and the
typically observed Gd moment of 8 $\mu_B$. Thus we can conclude
that all of the Gd ions in the sample are paramagnetic and coexist
with SC down to 1.8 K, despite the close proximity of the Gd ions
to the \cuo{} planes (1.7 \AA). This is evidence for a strong
confinement of the superconducting singlet spin pair wave function
to the \cuo{} planes. On the other hand, the c-axis
superconducting coherence length $\xi_c = 5.2$ \AA{} is longer
than the spacing between the Gd ion and the \cuo{} planes (1.7
\AA{}), and also exceeds the interplane distance, implying 3D SC.
The notion of 3D SC is corroborated by our STS studies,
\cite{Chen02} which probe the charge degrees of freedom and reveal
an isotropic s-wave superconducting gap. The apparent problem of
3D isotropic s-wave SC coexisting with strong Gd local moments
less than 1.7 \AA{} from the \cuo{} planes can be resolved by
considering the spin and charge (orbital) degrees of freedom of
the Cooper pairs separately. Whereas the singlet spin pairing is
confined to the \cuo{} planes, the orbital pair wave function
could still overlap adjacent \cuo{} planes, resulting in 3D SC for
all values of $T < T_c$ and $H < H_{c2}$. The bulk nature of the
SC is evident in the lower inset of Fig. \ref{SrGdCuO2}, which
shows the electronic contribution to $C$ plotted as $\Delta C/T$
vs $T$. The peak near 43 K is associated with \tc{}, and the ratio
$\Delta C/\gamma T_c$ is 2.9, assuming a Sommerfeld coefficient
\cite{Liu03} of $\gamma = 1.2$ mJ/mol K$^2$.

\epsfxsize=210pt

In conclusion, a large upper critical field anisotropy ratio
$\gamma = 8$ has been inferred from penetration depth measurements
of \slco{} in pulsed fields, despite the nearly cubic crystal
structure. The in plane upper critical field \hcab{} extrapolates
close to the Pauli paramagnetic limit $H_{c2}^P = 159$ T,
suggesting possible spin limiting for this orientation, as has
been observed in \YBCO{}. \cite{OBrien02} There is a large
separation between \hcab{} and the irreversibility field
\hirrab{}, which extends down to $T \rightarrow 0$, and is
abnormal for electron doped \hightc{} cuprates. This suggests the
existence of field-induced superconducting spin fluctuations
perhaps due to a competing SDW. In spite of the significant
anisotropy in the upper critical fields, $\xi_c$ is longer than
the spacing between \cuo{} planes, indicating three-dimensionality
of the orbital wave function. The low-field thermodynamic
measurements of $M(T)$ and $C(T)$ for \sgco{} polycrystals
indicate a coexistence of bulk SC with Gd paramagnetism, with the
full $J = 7/2$ Hund's rule moment despite the close proximity of
Gd atoms to the \cuo{} planes, and a $T_c$ of 43 K in both \sgco{}
and \slco{}. This can interpreted in terms of a strong confinement
of the spin degrees of freedom of the Cooper pairs to the \cuo{}
planes, whereas the orbital wave functions overlap adjacent \cuo{}
planes, and exhibit isotropic s-wave symmetry as determined by STS
measurements. \cite{Chen02}

This work was supported by the National Science Foundation under
Grant No. DMR-0103045 and DMR-0405088, and the National High
Magnetic Field Laboratory at Los Alamos, NM. V.Z. acknowledges
support by the Caltech Millikan Postdoctoral Fellowship program.

\newpage


\begin{thebibliography}{10}

\bibitem{Anderson87}
See, e.g. P. W. Anderson, ''The Theory of Superconductivity in
High-T$_c$
  cuprates," and references therein.

\bibitem{MapleBook}
M. B. Maple, E. D. Bauer, V. S. Zapf, and J. Wosnitza, 'Survey of
Important
  Experimental Results' in ''Unconventional Superconductivity in Novel
  Materials," Springer Verlag, to be published, and references therein.

\bibitem{Jung02a}
C.~U. Jung {\it et~al.}, Physica C {\bf 366},  299  (2002).

\bibitem{Liu01}
R.~S. Liu {\it et~al.}, Solid State Comm. {\bf 118},  367  (2001).

\bibitem{Kim02a}
M.-S. Kim {\it et~al.}, Phys. Rev. B {\bf 66},  214509  (2002).

\bibitem{Chen02}
C.-T. Chen {\it et~al.}, Phys. Rev. Lett. {\bf 88},  227002
(2002).

\bibitem{Jung02b}
C.~U. Jung {\it et~al.}, Phys. Rev. B {\bf 65},  172501  (2002).

\bibitem{Liu03}
Z.~Y. Liu {\it et~al.}, to appear in Europhys. Lett. (2004);
available at
  cond-mat/0306238 (unpublished).

\bibitem{Williams02}
G.~V.~M. Williams {\it et~al.}, Phys. Rev. B {\bf 65},  224520
(2002).

\bibitem{Hao91}
Z. Hao {\it et~al.}, Phys. Rev. B {\bf 43},  2844  (1991).

\bibitem{Kim02b}
M.-S. Kim {\it et~al.}, Solid State Comm. {\bf 123},  17  (2002).

\bibitem{Farrell89}
D.~E. Farrell {\it et~al.}, Phys. Rev. Lett {\bf 63},  782
(1989).

\bibitem{Farrell88}
D.~E. Farrell {\it et~al.}, Phys. Rev. Lett. {\bf 61},  2805
(1988).

\bibitem{Jung03}
C.~U. Jung, J.~Y. Kim, S.-I. Lee, and M.-S. Kim, Physica C {\bf
391},  319
  (2003).

\bibitem{Mielke01}
C. Mielke {\it et~al.}, J. Phys.: Condens. Matter {\bf 13},  8325
(2001).

\bibitem{Werthamer66}
N.~R. Werthamer, E. Helfand, and P.~C. Hohenberg, Phys. Rev. B
{\bf 147},  295
  (1966).

\bibitem{OBrien02}
J.~L. OBrien {\it et~al.}, Phys. Rev. B {\bf 61},  1584  (2000).

\bibitem{Fournier03}
P. Fournier and R.~L. Greene, Phys. Rev. B {\bf 68},  094507
(2003).

\bibitem{Demler01}
E. Demler, S. Sachdev, and Y. Zhang, Phys. Rev. Lett. {\bf 87},
067202
  (2001).

\bibitem{Sachdev02}
S. Sachdev and S.~C. Zhang, Science {\bf 295},  452  (2002).

\bibitem{Chen03}
C.-T. Chen and N.-C. Yeh, Phys. Rev. B {\bf 68},  220505(R)
(2003).

\bibitem{Chen04}
C.-T. Chen {\it et~al.} (unpublished).

\bibitem{Yeh04}
N.~C. Yeh {\it et~al.}, to appear in Physica C (2004).

\bibitem{Lee99}
Y.~S. Lee {\it et~al.}, Phys. Rev. B {\bf 60},  3643  (1999).

\bibitem{Kang03}
H.~J. Kang {\it et~al.}, Nature {\bf 423},  522  (2003).

\bibitem{Matsuura03}
M. Matsuura {\it et~al.}, Phys. Rev. B {\bf 68},  144503  (2003).

\bibitem{Lake02}
B. Lake {\it et~al.}, Nature {\bf 415},  299  (2002).

\bibitem{Mook02}
H.~A. Mook {\it et~al.}, Phys. Rev. B {\bf 66},  144513  (2002).

\end{thebibliography}
\end{document}